\documentclass{jfm}

\usepackage{graphicx}
\usepackage{amsmath}
\usepackage{amssymb}
\usepackage{natbib}
\usepackage{upmath}
\usepackage{bm}
\usepackage{draftcopy}
\usepackage{color}
\usepackage{subfig}

\usepackage[normalem]{ulem}

\providecommand\ui{\mathrm{i}}
\providecommand\bnabla{\boldsymbol{\nabla}}
\providecommand\bcdot{\boldsymbol{\cdot}}

\newcommand\be{\boldsymbol{e}}

\newcommand\br{\boldsymbol{r}}
\newcommand\bs{\boldsymbol{s}}
\newcommand\bu{\boldsymbol{u}}
\newcommand\bw{\boldsymbol{w}}
\newcommand\bv{\boldsymbol{v}}
\newcommand\bx{\boldsymbol{x}}
\newcommand\bz{\boldsymbol{z}}
\newcommand\bA{\boldsymbol{A}}
\newcommand\bF{\boldsymbol{F}}
\newcommand\bG{\boldsymbol{G}}
\newcommand\bH{\boldsymbol{H}}
\newcommand\bU{\boldsymbol{U}}
\newcommand\bzero{\boldsymbol{0}}
\newcommand\bth{\boldsymbol{\theta}}
\newcommand\bph{\boldsymbol{\phi}}
\newcommand\bps{\boldsymbol{\psi}}

\newcommand\cC{\mathcal{C}}
\newcommand\cP{\mathcal{P}}

\newcommand\Rey{\mbox{\textit{Re}}}  

\newcommand\eg{e.g.\ }
\newcommand\ie{i.e.\ }

\title[Patterns in Lipid Membrane Vesicles]{Shear-Driven Circulation Patterns in Lipid Membrane Vesicles}

\author[Francis G. Woodhouse and Raymond E. Goldstein]%
{F\ls R\ls A\ls N\ls C\ls I\ls S\ns G.\ns W\ls O\ls O\ls D\ls H\ls O\ls U\ls S\ls E \and 
R\ls A\ls Y\ls M\ls O\ls N\ls D \ns E.\ns G\ls O\ls L\ls D\ls S\ls T\ls E\ls I\ls N\thanks{Email address for correspondence: R.E.Goldstein@damtp.cam.ac.uk}
}
\affiliation{Department of Applied Mathematics and Theoretical Physics, 
University of Cambridge, Wilberforce Road, Cambridge CB3 0WA, UK}

\pubyear{2011}
\volume{???}
\pagerange{???--???}
\date{\today~ and in revised form ??}

\begin{document}

\maketitle

\begin{abstract}
Recent experiments have shown that when a near-hemispherical lipid vesicle attached to a solid
surface is subjected to a simple shear flow it exhibits a pattern of membrane circulation
much like a dipole vortex.  This is in marked contrast to the toroidal circulation that would
occur in the related problem of a drop of immiscible fluid attached to a surface and subjected 
to shear.  This profound difference in flow patterns arises from the lateral incompressibility 
of the membrane, which restricts the observable flows to those in which the velocity field in the 
membrane is two-dimensionally divergence free. Here we study these circulation patterns 
within the simplest model of membrane fluid dynamics.  A systematic
expansion of the flow field based on Papkovich--Neuber potentials is developed for general viscosity ratios
between the membrane and the surrounding fluids.  Comparison with experimental
results [C. V{\'e}zy, G. Massiera, and A. Viallat, {\it Soft Matter} {\bf 3}, 844 (2007)] is made,
and it is shown how such studies could allow measurements of the membrane viscosity.
Issues of symmetry-breaking and pattern selection are discussed.

\end{abstract}

\section{Introduction}
\label{sec:introduction}

One of the defining features of plant cells is the vacuole, an
organelle filled with water that acts as a nutrient storehouse for the cytoplasm.
The vacuole is contained within the vacuolar membrane, a lipid bilayer structure known also 
as the tonoplast.  It is a nearly ubiquitous feature of plants that the cytoplasm, the thin fluid layer surrounding
the vacuolar membrane, is in constant motion through the phenomenon of cytoplasmic 
streaming \citep{Lubicz2010} in which motor proteins move along filaments
and entrain fluid.  Dating back to important work by \citet{Pickard72} it has 
been suggested that the shear created in the cytoplasm is fully transmitted through the tonoplast
into the vacuolar fluid, a conjecture supported indirectly  by recent whole-cell 
measurements of the streaming velocity profile in large plant cells \citep{vandeMeent2010}.  

In considering the dynamics of shear transmission across the tonoplast,
ideas may be drawn from the large body of work on the response to flow of
lipid vesicles. These are closed bilayer membrane structures enclosing fluid, whose membrane
is generally in the `fluid' phase with zero shear modulus.
These studies constitute a highly active area of research complementing the
well-established understanding of membranes in thermal equilibrium \citep{Seifert1997}. 
The first insights into vesicle
motion under flow were made when \citet{Keller1982} described prototypical
`tank-treading' and `tumbling' behaviour. Further developments were quickly made in the
context of red blood cell dynamics; \citet{BarthesBiesel1985} used a visco-elastic membrane model to understand
the observed tank-treading behaviour, and \citet{Feng1989} provided detailed insights into the precise structure of the
area-preserving membrane flow patterns. However,
it is more recently that advances have been made in understanding the full three-dimensional
behaviour of nearly-spherical vesicles \citep{Seifert99}, revealing a rich phase space of both stable
\citep{Misbah06,Lebedev2007, Deschamps2009, Zhao2011} and unstable 
\citep{Kantsler2007} behaviours.

Less well-studied has been the hydrodynamics of vesicles when constrained to lie stuck to a surface.
Such problems arise as the natural extension of considering the flow of these objects near to walls, since
adhesion can occur upon the vesicle coming into contact with the surface \citep{Abkarian2008}.
Indeed, the elementary problem of an immiscible fluid droplet in contact with a no-slip plane
immersed in a bulk flow, absent the interfacial membrane of a vesicle, is one that has attracted detailed study.
\citet{Dussan1987} performed a comprehensive theoretical study of the behaviour of shallow droplets
in such situations, deriving a flow pattern whereby the droplet flows with the applied shear on the surface
and recirculates back underneath inside. \citet{Sugiyama2008} considered a perfectly hemispherical droplet
with similar conclusions. The presence of an interfacial membrane disallows such flows,
since they are not area-conserving and the membrane is unable to flow back through the vesicle,
and so a qualitatively different flow pattern is guaranteed simply by the inclusion of the membrane.

\citet{Lorz2000} studied the weak adhesion of vesicles
and observed a tank-treading type behaviour of the membrane under shear flow, with symmetric
recirculating regions on either side of the vesicle midline.
To further quantify the membrane hydrodynamics,
\citet{Vezy2007} studied adhered vesicles in shear flow with varying
contact areas, and observed the same symmetric doubly-recirculating pattern of motion of the membrane
in all cases. 
These experiments have yet to be complemented by any theoretical modelling and calculation of the expected membrane
and bulk flow patterns in such geometries. Therefore, as a first step towards
understanding these observations we formulate here a simplified theory for a vesicle adhered to a flat substrate
subjected
to shear flow. The vesicle is assumed to be an axisymmetric spherical cap, and the membrane to be
an impermeable, two-dimensional, incompressible Newtonian fluid, where we approximate the bilayer
structure of the membrane as a single layer by assuming zero inter-layer slip. On the phenomenological
length scales of this problem ($\sim 1-100\;\umu\text{m}$) the inter-layer dissipation is low \citep{Seifert1993}, and
so this simplification is justified. We will also assume that the vesicle adheres sufficiently strongly
to the surface so that we may neglect any shear-driven deformations allowed by remaining excess 
area.  (Typical experimental shear rates fall in the range $1-10\;\text{s}^{-1}$; see 
\citealt{Vezy2007}).  
Throughout we work in the Stokes regime $\Rey \ll 1$ for all fluids.

\section{Theory}

\subsection{Problem formulation}

\begin{figure}
\centering
\includegraphics{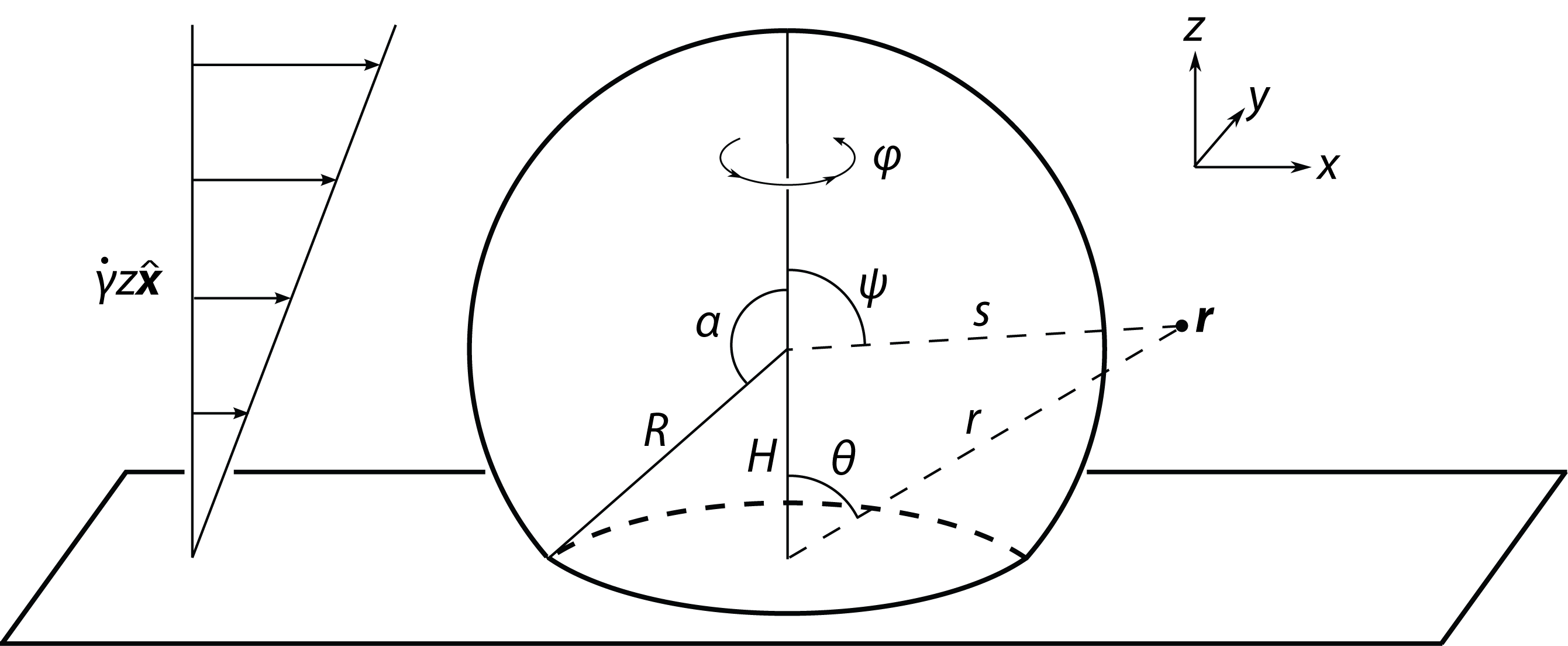}
  \caption{Geometry of the problem.}
\label{fig:coordinates}
\end{figure}

The vesicle is a spherical cap of radius $R$ and origin $x=y=0, z=H$, adhered to the plane $z=0$.
Let $(r,\theta,\phi)$ be spherical polars centred on $x=y=z=0$, and similarly let $(s,\psi,\phi)$ be centred on the origin of the adhering vesicle.
The vesicle then has a maximum extent $\psi=\alpha$ at $\theta=\upi/2$, where $\alpha = \upi - \arccos(H / R)$.
Figure \ref{fig:coordinates} illustrates the geometry.
We solve for three velocity fields: the flow $\bu^-$ inside the vesicle $s<R$,
the 2D flow $\bu^m$ on the membrane $s=R$, and the flow $\bu^+$ outside the vesicle $s>R$ which we will write as $\bu^+ = \bU^+ + \dot\gamma z\hat\bx$, where here and henceforth hatted quantities denote
coordinate system basis vectors.
To these fluids we associate respective viscosities $\eta_-$, $\eta_m$, $\eta_+$.

In what follows we non-dimensionalise
space by $R$, velocities by $R\dot\gamma$ and volume pressures by $\dot\gamma^2 \rho R^2$.
Boundary conditions are then applied on $s=1$ and the imposed shear rate is unity.
We also define $h \equiv H/R$, the effective height in this scaling.
The external flows obey the unforced Stokes and incompressibility equations,
\begin{align}
 \kappa_\pm \nabla^2 \bu^\pm - \bnabla p^\pm = \bzero, \qquad
 \bnabla \bcdot \bu^\pm = 0,
\end{align}
where $\kappa_\pm = \eta_\pm / \dot\gamma \rho R^2$.
We impose far-field asymptotics, planar no-slip and radial no-penetration:
\begin{align}
 \bu^+ \sim z\hat\bx & \text{ as } r \rightarrow \infty, \label{eq:bulk_asymptotic} \\
 \bu^\pm = \bzero & \text{ on } \theta=\upi/2, \label{eq:bulk_plane_noslip} \\
 \bu^\pm \bcdot \hat\bs = 0 & \text{ on } s=1. \label{eq:bulk_radial_nopen} 
\end{align}
All three velocities must be continuous across the membrane,
\begin{align}
\bu^+ = \bu^m = \bu^- & \text{ on } s=1, \label{eq:all_continuous}
\end{align}
which, along with \eqref{eq:bulk_plane_noslip}, implies the planar no-slip
condition $\bu^m = \bzero$ at $\psi=\alpha$.

In the absence of the membrane, we would only impose continuity of the bulk fluids' normal stresses 
at the interface, but since membranes can support tension, the bulk stresses may be discontinuous.  
By assumption the membrane satisfies the Stokes equations and incompressibility constrained to the surface $s=1$.
For a planar membrane this is simple \citep[\eg][]{Lubensky1996}, but curvature
induces an extra term. \citet{Henle2010} \citep[see also][]{Henle2008} formulate the Stokes equations in terms of
covariant derivatives
on an arbitrary manifold, from which the membrane tension
(equivalent to a two-dimensional pressure) may
be eliminated by taking the curl, as in the familiar derivation of the vorticity equation. 
In our notation, 
\begin{align}
 \hat\bs \bcdot \bnabla \times [ \hat\nabla^2 \bu^m + \bu^m ]
	+ \hat\bs \bcdot \bnabla \times \left[ \frac{2 \be^+_\parallel}{r_+} - \frac{2 \be^-_\parallel}{r_-} \right]_{s=1} = 0,
\quad
\hat\bnabla \bcdot \bu^m = 0, \label{eq:membrane_stressbc_noexpansion} 
\end{align}
where $\be^\pm_\parallel = e_{s \psi} \hat\bps + e_{s \phi} \hat\bph$ are the bulk fluids' in-plane normal
rates-of-strain,
$\hat\bnabla$ denotes the gradient operator constrained to the surface $s=1$, and we define
\begin{align}
 r_\pm \equiv \frac{\eta_m}{R \eta_\pm}.
\label{eq:sdlengths}
\end{align}
These are the non-dimensional form of the `Saffman-Delbr{\"u}ck' lengths $\ell_\pm \equiv \eta_m/\eta_\pm$
\citep{Saffman1975, Saffman1976, Henle2010}, and are the parameters with which we control the membrane dynamics.  Using the typical values
$\eta_\pm \sim 10^{-2}\;\text{Poise}$ (water),
$\eta_m \sim 10^{-7}-10^{-5}\;\text{Poise cm}$ \citep{Camley2010} and $R \sim 1-100\;\umu\text{m}$, 
we find relevant experimental ranges to be $r_\pm \sim 10^{-3}-10$.

\subsection{Solution method}
\label{sec:solution_method}

We will first construct a set of basis functions for the bulk fluids such that the planar no-slip condition \eqref{eq:bulk_plane_noslip}
is automatically satisfied. A common approach to solving Stokes flow problems is to use Lamb's solution \citep{Happel1991},
and \citet{Ozarkar2008} showed that it is possible to write down a no-slip image system for each individual Lamb mode
in the manner of the original solution of \citet{Blake1971} for a simple Stokeslet. However,
this becomes algebraically unwieldy, and we adopt a slightly different approach.

A method similar to using a Lamb expansion is to write down the solution in terms of Papkovich--Neuber potentials \citep{TranCong1982}:
if $\nabla^2 \bA=\bzero$ and $\nabla^2 B=0$,
\begin{alignat}{2}
 \bv & = \bnabla( \br \bcdot \bA + B ) - 2\bA,   &\quad   p & = 2\eta \bnabla \bcdot \bA,   \label{eq:pn_soln}
\end{alignat}
is a solution to the incompressible Stokes equations $\eta\nabla^2 \bv = \bnabla p$, $\bnabla \bcdot \bv = 0$. This reduces the
problem to one of solving the vector and scalar harmonic equations.
\citet{Shankar2005} uses this representation to solve for Stokes flow inside a circular cone using three harmonic basis functions,
one scalar and two vector. The scalar harmonic is the usual solution to Laplace's equation in spherical coordinates,
\begin{align}
 B(\br;\nu,m) = r^\nu e^{\ui m \phi} P^m_\nu(\cos \theta),
\end{align}
where $\nu \in \mathbb{R}$ (since the solution need not be analytic in $z<0$)
and $m \in \mathbb{Z}$, and $P^m_\nu$ are the generalised associated Legendre functions. Two independent vector harmonics can then
be derived from $B$ as $\bA_1 = \hat\br \times \bnabla(rB)$ and $\bA_2 = B \hat\bz$, which evaluate to
\begin{align}
 \bA_1(\br;\nu,m) &=  -r^\nu e^{\ui m \phi} [ m \csc \theta P^m_\nu(\cos \theta)\hat\bth + \ui {P^m_\nu}'(\cos \theta) \hat\bph ], \\
 \bA_2(\br;\nu,m) &=  r^\nu e^{\ui m \phi} [ \cos\theta P^m_\nu(\cos \theta)\hat\br - \sin\theta P^m_\nu(\cos \theta)\hat\bth ],
\end{align}
where primes indicate differentiation with respect to $\theta$.
Since the only non-zero imposed velocity is simple shear flow we only need the $m=1$ mode, and can take the real part of all
basis functions, whose $m$-dependence we henceforth omit. We also write $\cP_\nu(x) \equiv P^1_\nu(x)$.

The next step is to find combinations of these harmonics such that the no-slip condition \eqref{eq:bulk_plane_noslip} is satisfied.
In \citet{Shankar2005} this is done by evaluating the velocity field on the cone using one unit of $B(\br;\nu+1)$,
$a$ units of $\bA_1(\br;\nu)$ and $b$ units of $\bA_2(\br;\nu)$.
However, the case of a flat plane is a special one and solutions with no contribution from $B$ are possible,
so here we must take an arbitrary $c$ units of $B(\br;\nu+1)$; then \eqref{eq:pn_soln} gives
\begin{subequations}
\label{eq:v_cpts}
\begin{align}
 v_r &= r^\nu \cos\phi \left[ c(\nu+1)\cP_{\nu+1} + b(\nu-1)\cos\theta\,\cP_\nu \right], \label{eq:v_r_cpt} \\
 v_\theta &= r^\nu \cos\phi \left[c\cP'_{\nu+1} + 2a \csc \theta \,\cP_\nu + b\left\{\sin\theta\,\cP_\nu + 
\cos\theta\, \cP'_\nu\right\}\right], \label{eq:v_theta_cpt} \\
 v_\phi  &= -r^\nu \sin\phi \left[c \csc\theta\, \cP_{\nu+1} + 2a \cP'_\nu + b \cot \theta\, \cP_\nu\right]. \label{eq:v_phi_cpt}
\end{align}
\end{subequations}
(From here on, where unspecified the argument of $\cP_\nu$ is $\cos\theta$.)
It is simple to find that demanding $\bv=\bzero$ at $\theta=\upi/2$ leads to two possible conditions:
\begin{alignat}{4}
 \text{Either} &\quad& 2a+b-(\nu+2)c = 0 &\quad& \text{and} &\quad& \cP_{\nu+1}(0) &= 0, \\
 \text{or} && a=c=0 && \text{and} && \cP_\nu(0) &= 0.
\end{alignat}
The condition $\cP_\nu(0) = 0$ has solutions $\nu = 0,2,4,\ldots$ and $\nu=-1,-3,-5,\ldots$.
(Note that $\cP_0 \equiv \cP_{-1} \equiv 0$.)

With this basis the general solution with no-slip on $z=0$ can be written
\begin{align}
 \bv(r,\theta,\phi) = \sum_{\nu=-\infty}^\infty r^\nu [a_\nu \bF_\nu(\theta,\phi) + b_\nu \bG_\nu(\theta,\phi) + c_\nu \bH_\nu(\theta,\phi)],
\end{align}
where the functions $\bF,\bG,\bH$ may be read off from \eqref{eq:v_cpts} as the coefficients of $a,b,c$ respectively. The coefficients must satisfy
\begin{alignat}{3}
 2a_\nu + b_\nu - (\nu+2)c_\nu = 0 && \text{ for } && \nu &= 1,3,5,\ldots, \nu = -2,-4,-6,\ldots, \\
 a_\nu = c_\nu = 0 &\,& \text{ for } &\,& \nu &= 2,4,6,\ldots, \nu = -3,-5,-7,\ldots.
\end{alignat}
These constraints can be incorporated directly into the expansion by writing
\begin{align}
 \bv(r,\theta,\phi) = \sum_{\substack{\nu=2,4,6,\ldots \\ \nu=-3,-5,\ldots}} r^\nu b_\nu \bG_\nu(\theta,\phi)
			+ \sum_{\substack{\nu=1,3,5,\ldots \\ \nu=-2,-4,\ldots}} r^\nu [ a_\nu \bF^{(1)}_\nu(\theta,\phi) + c_\nu \bF^{(2)}_\nu(\theta,\phi) ]
\label{eq:v_full_expansion}
\end{align}
where $\bF^{(1)}_\nu = \bF_\nu - 2\bG_\nu$ and $\bF^{(2)}_\nu = \bH_\nu + (\nu+2)\bG_\nu$.

We now expand $\bu^-$ and $\bU^+$ in this fashion. Requiring regularity of $\bu^-$ at $r=0$ implies that $a^-_\nu,b^-_\nu,c^-_\nu$
are non-zero only for $\nu > 0$. Similarly, for $\bu^+ = z\hat\bx + \bU^+$ to satisfy the asymptotic condition \eqref{eq:bulk_asymptotic}
we must have $\bU^+ \rightarrow 0$ as $r \rightarrow \infty$ and so $a^+_\nu,b^+_\nu,c^+_\nu$ are non-zero only for $\nu<0$.

We now turn our attention to the flow on the membrane surface $s=1$. In \citet{Henle2010} it is shown that the surface flow $\bu^m(\br)$
can be decomposed into a sum of incompressible shear modes $\bw(\psi,\phi;\mu) = \bs \times \bnabla \Phi(\psi,\phi;\mu)$
where the scalar functions $\Phi(\psi,\phi;\mu)$ are eigenfunctions of the Laplacian, $\nabla^2 \Phi(\psi,\phi;\mu) = -\mu(\mu+1) \Phi(\psi,\phi;\mu)$.
For the same reasons as in the bulk flow, we may take $\Phi(\psi,\phi;\mu) = \cos\phi \cP_\mu(\cos \psi)$, whence
\begin{align}
 \bw(\psi,\phi;\mu)  =  \frac{2}{\sin \psi} \cP_{\mu}(\cos \psi) \cos \phi \,\hat\bps -  2 \cP'_{\mu}(\cos \psi) \sin \phi \,\hat\bph.
\end{align}
Given a mode spectrum $\{\mu_n\}$, expand the velocity as $\bu^m(\psi,\phi) = \sum_n A_n 
\bw(\psi,\phi;\mu_n)$.
Incompressibility is satisfied for each mode by construction. Then the membrane dynamics \eqref{eq:membrane_stressbc_noexpansion} become
\citep{Henle2010}
\begin{align}
 \sum_n A_n [2 - \mu_n(\mu_n+1)] \mu_n(\mu_n+1) \Phi(\psi,\phi;\mu_n)
    + \hat\bs \bcdot \bnabla \times \left[ \frac{2 \be^+_\parallel}{r_+} - \frac{2 \be^-_\parallel}{r_-} \right]_{s=1} = 0. \label{eq:membrane_stressbc}
\end{align}

\begin{figure}
\centering
\includegraphics{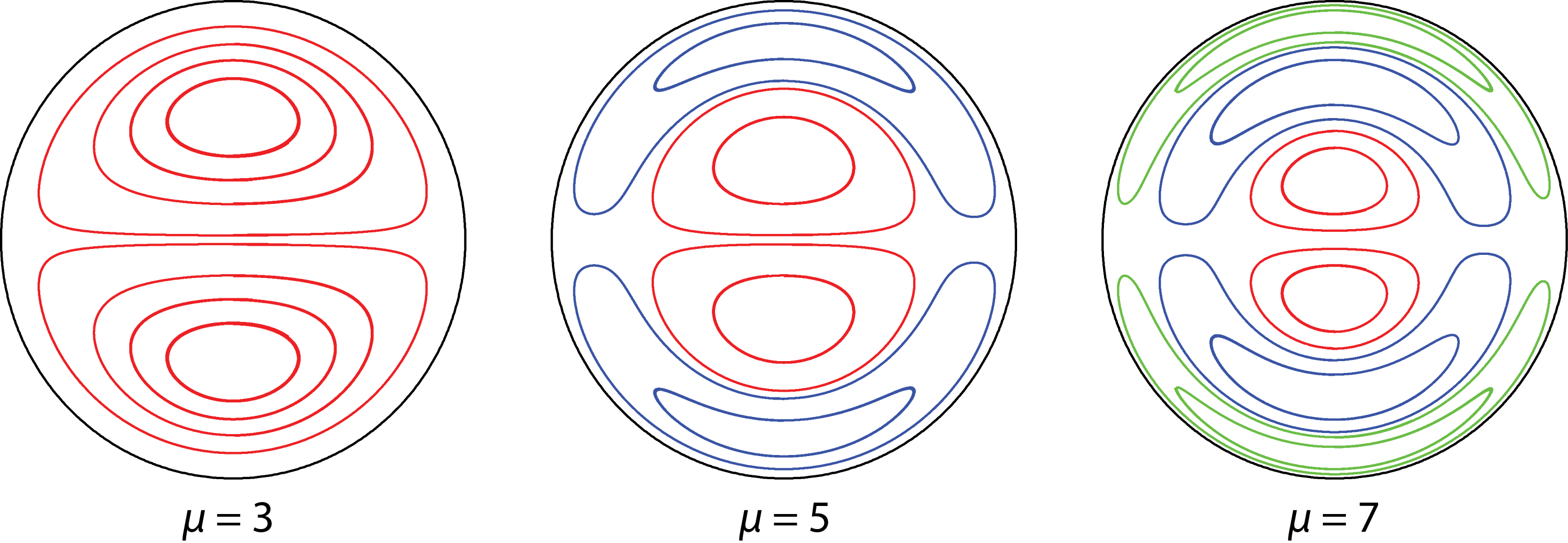}
 \caption{Streamlines of recirculating modes for $h=0$, projected onto the $z$-plane.}
\label{fig:possible_surface_flows}
\end{figure}

There still remains the question of which modes $\mu$ to sum over. The ideal basis set would satisfy no-slip at $\psi=\alpha$ in the same way as
the bulk fluid basis functions. For the membrane basis there is not enough freedom, so we can only choose
one component to automatically be zero. We choose to set $u^m_\psi(\alpha,\phi;\mu)=0$ for all $\phi$, which implies
\begin{align}
 0 = \cP'_\mu(\cos \alpha) = (1+\mu) \cot\alpha \cP_\mu(\cos \alpha) - \mu \csc\alpha  \cP_{\mu+1}(\cos \alpha).
\end{align}
We can also take $\mu > 0$ by the identity $\cP_{-(\mu+1)} \equiv \cP_\mu$.
For the case $\alpha=\upi/2$, \ie a perfect hemisphere ($h=0$), this equation has exact solutions $\mu = 1,3,5,\ldots$; if
$\alpha > \upi/2$, this must be solved numerically by a method such as Newton--Raphson to obtain the spectrum of allowed modes.
Figure \ref{fig:possible_surface_flows} shows streamlines of three different recirculatory modes for the hemispherical case,
demonstrating the increase in number of circulation centres as the mode number increases. The same qualitative patterns
persist for $h>0$.

We now have seven sets of coefficients to solve for in the expansions: $a^\pm_\nu, b^\pm_\nu, c^\pm_\nu$ for $\bu^\pm$,
and $A_n$ for $\bu^m$. These will be determined by satisfying the simple boundary
conditions \eqref{eq:bulk_radial_nopen} and \eqref{eq:all_continuous}, and the stress
boundary condition \eqref{eq:membrane_stressbc}. Each component of each boundary condition will only have $\phi$ dependence
in the form of an overall factor of $\sin \phi$ or $\cos \phi$, so we may factor these out and need only
satisfy boundary conditions for the $\psi$-dependent factors.

The complexity of the system does not lend itself easily to significant analytic progress, and so we choose to
proceed via numerical methods to determine the coefficients. Our chosen solution method is the least-squares
error minimisation procedure as outlined by \citet{Shankar2005}.
The velocity expansions are truncated at orders $N_1$ and $N_2$ for the bulk and membrane velocities respectively --
note that use of least-squares minimisation is crucial here, since any attempt to solve exactly for a finite
set of coefficients will yield poorly-determined systems.
The boundary conditions are to be satisfied on $0 \leq \psi \leq \alpha$. Divide the interval $[\epsilon,\alpha]$ ($\epsilon \ll 1$)
into $M$ equal subintervals, separated by points $\{\psi_k\}$. For each boundary condition $\cC_i$ to be enforced,
write $\varepsilon_{ik}$ for the error in satisfying that condition at $\psi_k$. Minimising $E^2 = \sum_{i,k} \varepsilon_{ik}^2$
with respect to $a^\pm_\nu, b^\pm_\nu, c^\pm_\nu, A_n$ then gives a linear system which can be solved for the
coefficients. As $N_1,N_2,M \rightarrow \infty$ the coefficients quickly approach limiting values. (In what
follows, we typically use magnitudes $N_1,N_2,M \sim O(100)$.)
Care must be taken in satisfying the stress condition \eqref{eq:membrane_stressbc}, since the stress becomes divergent
in the neighbourhood
of $\psi=\alpha$ due to the discontinuous geometry. To avoid this we truncate the collocation points $\{\psi_k\}$ a small
distance before $\psi=\alpha$ for this condition.

Due to the choice of origin for the bulk fluid expansion, the system quickly becomes numerically unstable as $h$ increases
due to divergences in $r^\nu$ for $\nu>0,r>1$ and $\nu<0,r<1$ in evaluating the velocity expansions on $s=1$. However, if the bulk
velocity were expanded about the origin of the vesicle we would no longer have a basis set satisfying planar no-slip, and would
have to enforce boundary conditions along the entire $y=z=0$ line as well, which would also possess highly divergent coefficients.

\section{Results and discussion}
\label{sec:results}

\begin{figure}
\centering
\includegraphics{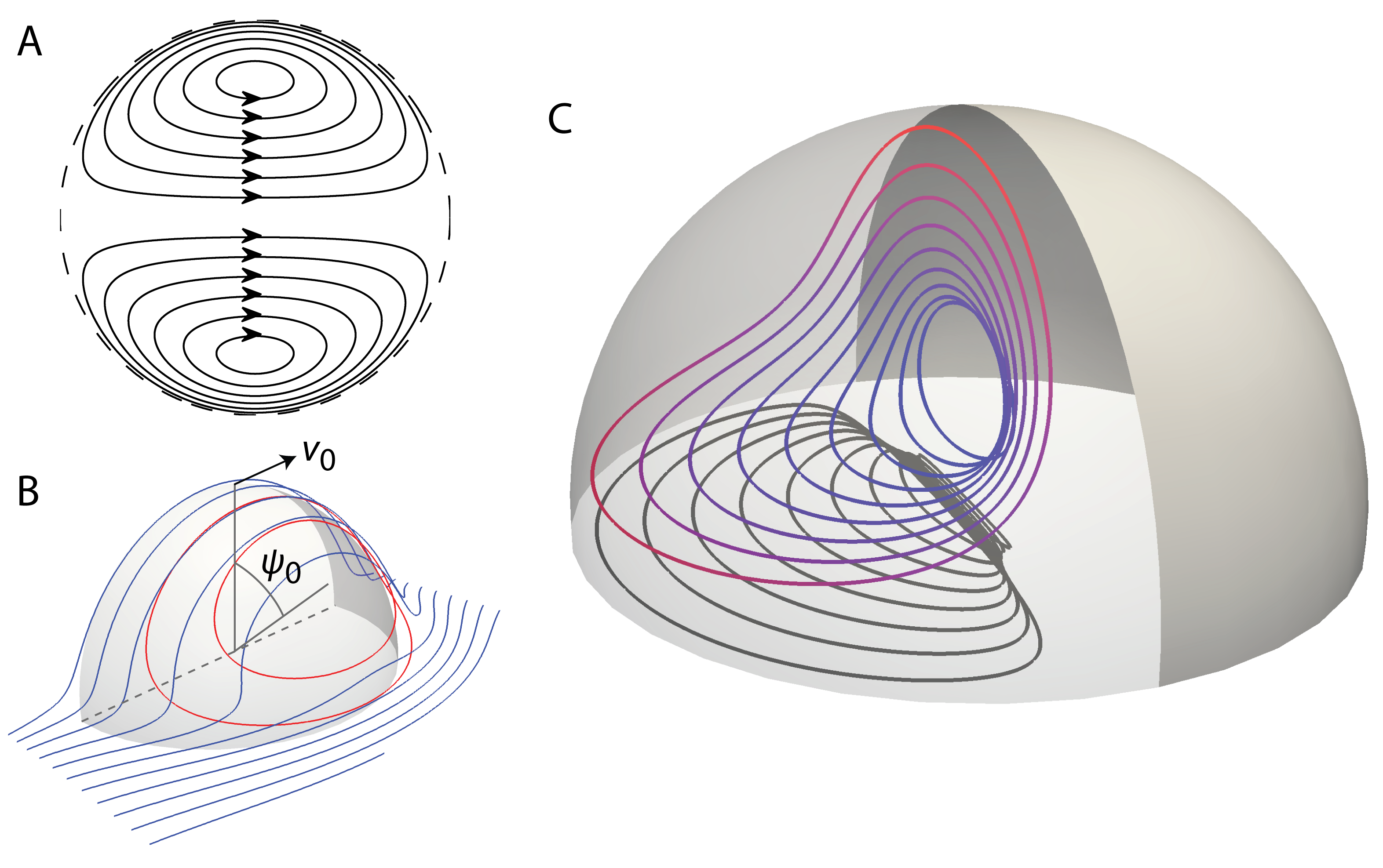}
\caption{Streamlines for $r_+ = r_- = 0.01, h = 0$. (a) Top view of membrane. (b) External flow,
with closed streamlines in red. (c) Internal flow; brighter red indicates faster flow.}
 \label{fig:flow_streamlines}
\end{figure}

Figure \ref{fig:flow_streamlines}a illustrates the typical flow induced on the surface of the vesicle. The closed,
two-lobed symmetric recirculating
patterns as observed by \citet{Vezy2007} are clearly reproduced, indicating a dynamical preference for
motion rather than remaining stationary. The same structure persists for non-zero values of $h$ and for varying viscosity ratios $r_\pm$.
The recirculatory flow of the membrane induces a flow inside the vesicle appearing to possess entirely closed
streamlines, as illustrated in figure \ref{fig:flow_streamlines}c. Similarly, the external flow
shows regions of recirculation on either side of the membrane which also appear
to contain closed streamlines (figure \ref{fig:flow_streamlines}b).

The qualitative difference seen between this problem and that of an immiscible hemispherical droplet
\citep{Sugiyama2008} is due to the presence of the membrane. As described in~\S\ref{sec:introduction},
simple droplet flow permits both a surface-compressible structure and mass exchange with the interior. Due to
the membrane structure, adhesion and incompressibility here,
no such globally circulatory flow is possible.

The dynamical insistence on motion can be understood by the following argument. Consider the related problem of
a free, spherical vesicle in uniform, unidirectional flow, for which it is known no membrane motion occurs.
This problem is axisymmetric about the flow direction,
and so the membrane experiences a tension constant along `lines of latitude' perpendicular to the flow direction.
Such a tension has no shearing component, and so the membrane is able to remain stationary. When the symmetry
is broken by the introduction of the adhesive wall and the application of shear flow, the tension
on the membrane is no longer uniform across `lines of latitude'. The membrane is unable to support this shear,
forcing it to flow. The same symmetry-breaking effects have been observed in vesicle membrane flows driven
by electric fields \citep{Staykova2008} where a small field inhomogeneity develops
shear stresses, causing the membrane to flow.

It should be noted that many surface-incompressible patterns are possible whilst remaining consistent with
the symmetries of the problem; figure \ref{fig:possible_surface_flows} shows the first three recirculatory
modes of the membrane flow basis constructed in~\S\,\ref{sec:solution_method}. We observe that the flow adopts a pattern
containing only two circulation centres rather than four or more, and pose the explanation 
that this is a consequence of dissipation reduction: the membrane is adopting the fewest number of topological `defects'
while still minimising dissipation within the bulk fluids.

\begin{figure}
\centering
\includegraphics{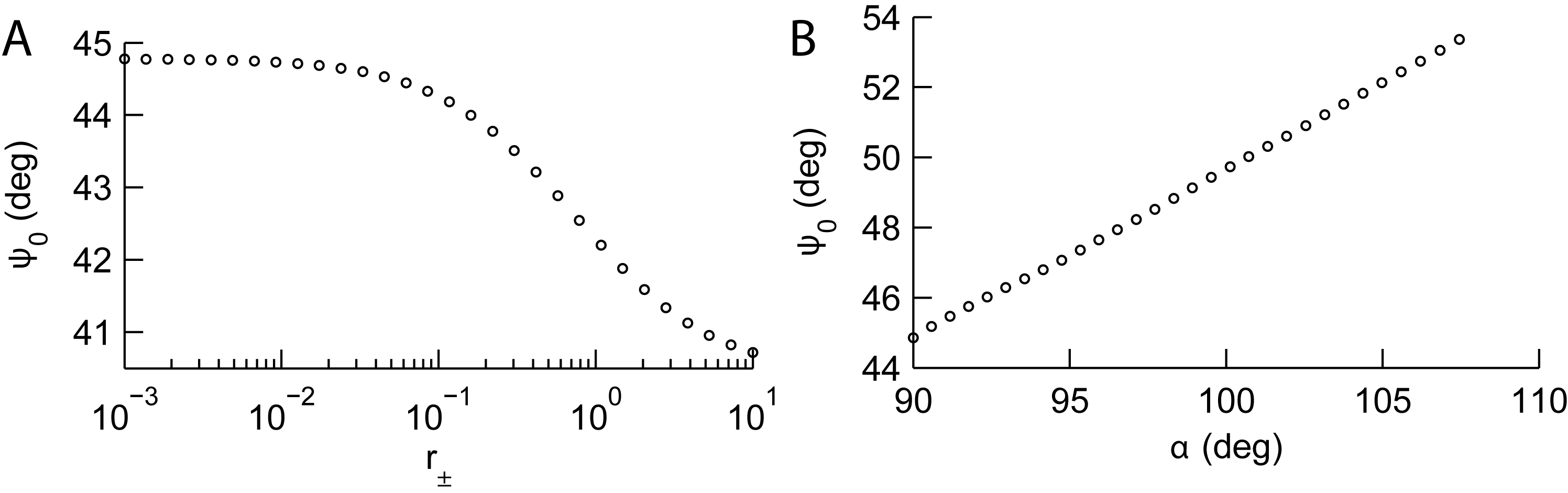}
 \caption{Dependence of the surface stagnation point angle on $r_\pm$ and $\alpha$.
(a) $\alpha=\frac{\pi}{2}$, $r_+=r_-$, $10^{-3} \leq r_\pm \leq 10$, log scale in $r_\pm$. 
(b) $r_+=r_-=10^{-2}$, $0<h<0.3$.}
\label{fig:stagnation_shift}
\end{figure}

As the membrane viscosity is increased through the physical range, the angle $\psi_0$ of the stagnation point from the
vertical (see figure \ref{fig:flow_streamlines}b)
on either side shifts downwards, illustrated in figure \ref{fig:stagnation_shift}a. The locations seen here are close to
those
observed by \citet{Vezy2007} (where their variable $\theta$ is equivalent to our $\frac{\pi}{2}-\psi$) for small values of $h$
(their $L/2R$ close to unity). We also observe that for $0<h<0.3$ the stagnation angle appears to scale
linearly with the base angle $\alpha$, as might be expected (figure \ref{fig:stagnation_shift}b).

\begin{figure}
\centering
\includegraphics{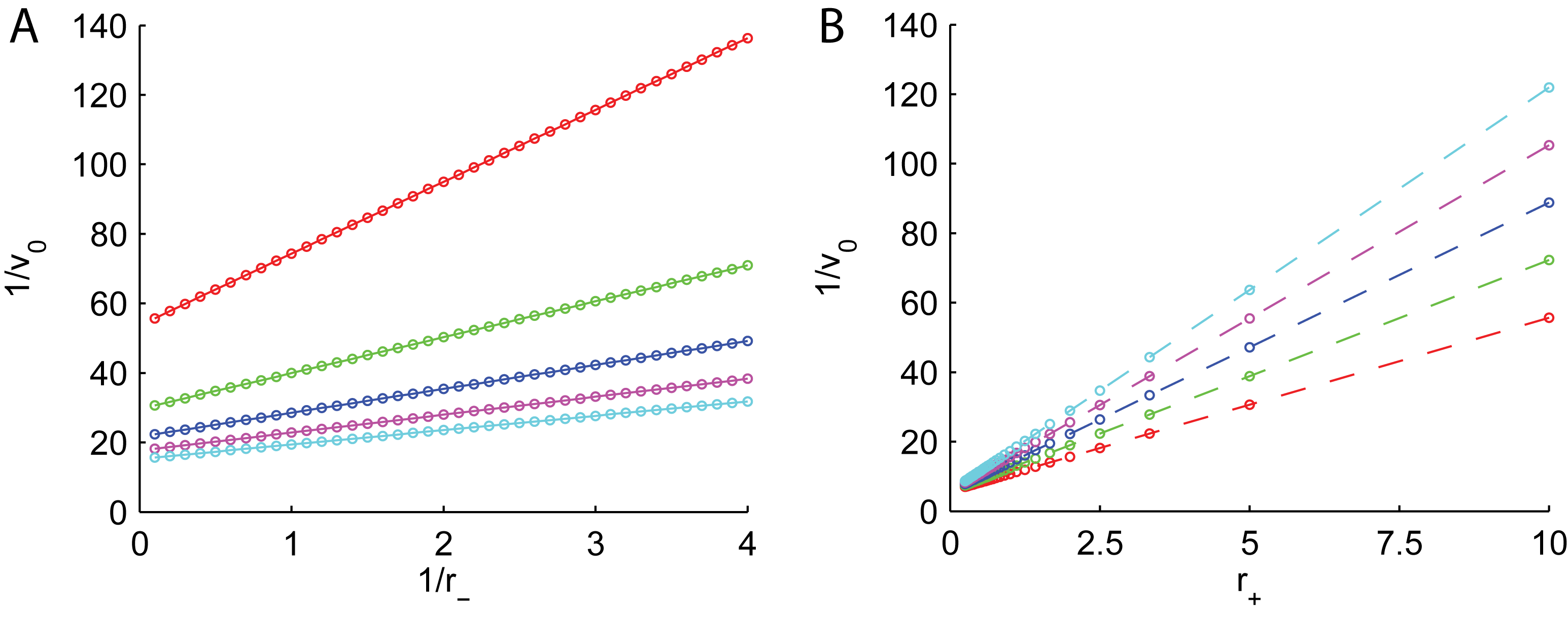}
 \caption{Linear dependence of $1/v_0$ on $1/r_-$ and $r_+$, at fixed values of $r_+$ and $r_-$, respectively. Top to
bottom, (a) $r_+ = 10,5,3.33,2.5,2$, (b) $r_- = 0.3,0.4,0.59,1.11,10$.}
\label{fig:membrane_viscosity_scaling}
\end{figure}

We now turn to the problem of experimentally
determining the membrane viscosity $\eta_m$. Due to the order of magnitude of the physical
constants involved, this is a difficult measurement to achieve. \citet{Dimova1999} proposed a method
based upon measuring the fluctuations of a spherical particle stuck on or penetrating through a vesicle's surface. While this method has been experimentally verified, 
and is perhaps simpler to initiate than 
encouraging vesicles to strongly adhere to substrates,
it requires difficult measurements of microscopic parameters.
An added benefit of the calculations presented here is a possible
complementary method of membrane viscosity measurement based solely on the vesicle flow profiles,
provided that the vesicle itself has not been affected by the adhesion process.
Such velocities can be measured either by the defect-tracking method of
\citet{Vezy2007} or by tracking small phase-separated patches of a second lipid species.
Figure \ref{fig:membrane_viscosity_scaling} suggests a fit of the apex velocity
$v_0 \equiv | \bv^m(\psi=0) |$ (see figure \ref{fig:flow_streamlines}b) with $r_\pm$ of the form
\begin{align}
 v_0 = \frac{r_-}{Ar_+ + Br_+ r_- + Cr_-}
\end{align}
for a hemispherical membrane ($h=0$). Therefore given known bulk viscosities
$\eta_\pm$, vesicle radius $R$ and measured apex velocity $v_0$, it would be possible to solve the above relationship
and so predict the membrane viscosity $\eta_m$.
Using $40\times 40$ regularly-spaced datapoints $(r_+,r_-) \in [0.25,10]^2$
we find an excellent fit for parameter values $A = 2.054$, $B = 4.718$, $C = 5.803$,
with root mean square error on the order of $10^{-4}$ and 95\% confidence intervals $\sim\pm0.001$.
We expect such a relationship to persist for $h \neq 0$.

\section{Conclusions}
\label{sec:conc}

We have effected the theoretical solution of the problem of near-hemispherical lipid vesicles adhered
to a flat substrate as experimentally investigated by \citet{Vezy2007}. We demonstrated the presence
of symmetric recirculating regions on the membrane and explained the qualitative difference
from the membrane-less case. Additionally, the existence
of closed streamlines both inside and outside the vesicle was shown.
We then used knowledge of this system to suggest a new method for estimating membrane viscosities.

Throughout this work we have assumed the membrane remains spherical. For low shear rates, this
assumption is valid; however, behaviour at higher shear rates is unlikely to remain
stable. \citet{Li1996} numerically studied the deformation of adhered immiscible liquid drops
in shear flow and found a regime where the droplet would continue to deform over time.
It is reasonable to suggest such deformations could occur in the case studied here,
though the more complex surface and internal flow structure may lead to a different
parameter phase space of deformations and instability.

Understanding the membrane flow has implications for other lipid membrane systems.
One particular area where knowledge of the hydrodynamic behaviour of attached vesicles
under shear flow is of importance
is in the study of transient vesicle adhesion. If a vesicle loaded with binders comes into
contact with a surface to which it may adhere, then the evolution of the contact area over
time is dependent on the forcing experienced by the vesicle.
\citet{BrochardWyart2002} developed a dynamical adhesion force model based on
the diffusion of binders on the membrane. However, if the vesicle is subjected to shear
then free binders will be advected by the membrane flow which may significantly
enhance or impair adhesive effects.

Possible future work in this area includes a more complex numerical study
in order to accommodate all sizes of spherical cap (i.e. larger $h$),
and experimental studies using the proposition of~\S\ref{sec:results}
to estimate membrane viscosities. It would also be interesting to study
shear rates higher than those in \citet{Vezy2007} to understand whether buckling
and deformation occur.

It is a pleasure to offer this contribution in honour of Tim Pedley's 70th birthday.
We thank P. Khuc Trong, P. Olla and J. Dunkel for helpful discussions and S. Ganguly for insights 
at an early stage of this
work. This work was
supported in part by the EPSRC and the European Research Council Advanced Investigator Grant 247333.

\bibliographystyle{jfm}
\bibliography{vesicleshear}{}

\end{document}